\documentclass[12pt,a4paper,final]{iopart}

\usepackage{graphicx}
\expandafter\let\csname equation*\endcsname\relax
\expandafter\let\csname endequation*\endcsname\relax
\usepackage{amsmath,amsthm,amsfonts}

\usepackage[breaklinks=true,colorlinks=true,linkcolor=blue,urlcolor=blue,citecolor=blue]{hyperref}

\newtheorem{example}{Example}

\usepackage{xcolor}

\begin{document}

\title[Combinatorial origins of the canonical ensemble]{Combinatorial origins of the canonical ensemble}

\author{Kornelia Ufniarz$^1$}
\ead{kornelia.ufniarz@gmail.com}
\author{Grzegorz Siudem$^1$}
\ead{grzegorz.siudem@pw.edu.pl}

\address{$^1$Faculty of Physics, Warsaw University of Technology,\\ Koszykowa 75, 00-662 Warsaw, Poland}

\begin{abstract}
The Darwin-Fowler method in combination with the steepest descent approach is a common tool in the asymptotic description of many models arising from statistical physics. In this work, we focus rather on the non-asymptotic behavior of the Darwin-Fowler procedure.  By using a combinatorial approach based on Bell polynomials, we solve it exactly. Due to that approach, we also show relationships of typical models with combinatorial Lah and Stirling numbers.
\end{abstract}

\vspace{2pc}

\section{Introduction}

Proposed by Darwin and Fowler \cite{Darwin1922a,Darwin1922b} method of steepest descent is a typical approach for the derivation of canonical (Gibbs) ensembles (see \cite{Fowler1952,Huang1978,Pathria2011}) or other asymptotic problems on the border on combinatorics and statistical physics \cite{Flajolet2009}. The method bases on the properties of the complex integrals and allows one to efficiently calculate the desired limit of the typical combinatorial problems. Thought, typically Darwin-Fowler approach is concerned with the asymptotics, in this article, we focus mainly on the non-asymptotic case. With the introduced combinatorial approach based on Bell polynomials we solve exactly the Darwin-Fowler problem purely combinatorically without integral representation.

 Bell Polynomials have been applied recently to the wide range of problems of statistical physics e.g. in the description of gas of clusters \cite{AFronczak2013,Siudem2013}, partition function for ideal gases \cite{Zhou2018}, series expansion for quantum partition functions \cite{Hen2018}, general lattice models description \cite{AFronczak2014} and Ising model \cite{Siudem2014}. With this work, we complete this list with the application to the canonical ensemble, which additionally reveals unexpected combinatorial origins of this fundamental to statistical mechanics distribution. With the introduced approach we show that for the typical  degeneracies (i.e. constant one and harmonic oscillator's) the most probable configurations are given by the well-known combinatorial Lah and Stirlings numbers.

The paper is composed as follows: In the first section, we introduce the classic formulation of the Darwin-Fowler method. The second section focuses on the definition and basic properties of Bell polynomials and their relationship to Lah and Stirling  numbers. In the third section, we use the derived approach to solving the Darwin-Fowler procedure for the non-asymptotic case.
The work is concluded with the analysis of the few examples of the degeneracy of energy levels, which reveals their relationship with the   Lah and Stirling numbers.

\section{Darwin-Fowler method}
Let us consider an ensemble consisting of the total of $\mathcal{K}$ systems with possible energies $E=\{\varepsilon_1,\,\varepsilon_2,\,\dots\}$ with degeneracies $\Omega=\{\omega_1,\, \omega_2,\,\dots\}$, where the energy level $\varepsilon_\ell$ has degeneracy equal to $\omega_\ell$. Furthermore, we assume that there exists a~quantum of the energy $\varepsilon$, which without loss of generality means that $\varepsilon_\ell= \varepsilon\ell$. The above allows one to formulate the following condition for $\{n_\ell\}$ i.e. the numbers of systems with the energies from $E$
\begin{equation}\label{eq:cond}
    \begin{cases}
    \sum_{\ell=1}^{\infty}n_\ell=\mathcal{K},\\
	\sum_{\ell=1}^{\infty} \varepsilon_\ell n_\ell=\varepsilon\mathcal{N}.
    \end{cases}
    \Longrightarrow\quad\;
        \begin{cases}
    \sum_{\ell=1}^{\infty}n_\ell=\mathcal{K},\\
	\sum_{\ell=1}^{\infty}\ell n_\ell =\mathcal{N}.
    \end{cases}
\end{equation}
In the Darwin-Fowler method we examine the probability distribution over the energy levels
\begin{equation}\label{eq:P-def}
    \mathcal{P}_{\mathcal{N},\,\mathcal{K}}(\ell)=\frac{n^\star_{\mathcal{N},\,\mathcal{K}}(\ell)}{\mathcal{K}},
\end{equation}
where $n^\star_{\mathcal{N},\,\mathcal{K}}(\ell)$ is the number of the systems at $\ell$-th energy level for the average configuration of ensemble consisting of $\mathcal{K}$ systems and having total energy $\varepsilon \mathcal{N}$. We assume that every single configuration occurs with equal probability (with respect to the degeneracies) thus
\begin{equation}\label{eq:oczekiwana}
	n^\star_{\mathcal{N},\,\mathcal{K}} = \frac{\sum_{\{m_r\}_{\mathcal{N},\,\mathcal{K}}}m_\ell W_{\mathcal{N},\,\mathcal{K}}(\{m_r\})}{\sum_{\{m_r\}_{\mathcal{N},\,\mathcal{K}}} W_{\mathcal{N},\,\mathcal{K}}(\{m_r\})},\quad\mathrm{where}\quad	W_{\mathcal{N},\,\mathcal{K}}(\{m_\ell\})=\mathcal{K}!\prod_{\ell}\frac{\omega_\ell^{m_\ell}}{m_\ell!},
\end{equation}
and the summation is taken over all possible configurations satisfying Eq.  (\ref{eq:cond}) and factor $W_{\mathcal{N},\,\mathcal{K}}(\{m_\ell\})$ counts the number of possible realisations of the configuration $\{m_\ell\}$. With the introduced notion one can realize that for the function $\Gamma_{\mathcal{N},\,\mathcal{K}}$ defined as 
\begin{equation}\label{eq:Gamma}
	\Gamma_{\mathcal{N},\,\mathcal{K}}(\omega_1,\,\omega_2,\,\dots)=\mathcal{K}!\sum_{\{m_r\}_{\mathcal{N},\,\mathcal{K}}}\left(\frac{\omega_1^{m_1}}{m_1!}\cdot \frac{\omega_2^{m_2}}{m_2!} \cdots\right),
\end{equation}
one can express Eq. (\ref{eq:oczekiwana}) for  the number $n^\star_{\mathcal{N},\,\mathcal{K}}$  in the following way
\begin{equation}
n^\star_{\mathcal{N},\,\mathcal{K}}=\omega_\ell\frac{\partial \ln\Gamma(u_1,\,u_2,\,\dots)}{\partial u_\ell}\Biggr|_{u_r=\omega_r,\,\,r=1,\,2,\,\dots}.\label{eq:wartoscocz}
\end{equation}

As we have noticed, the problem of determining the $P_{\mathcal{N},\,\mathcal{K}}$ distribution can be reduced to determining the value of the function $\Gamma_{\mathcal{N},\,\mathcal{K}}$ and its derivative. Typically, in the Darwin-Fowler method, this is done by using the steepest descend method (see \ref{sec:steepestdescend}). Instead of this approximate approach, we solve the problem exactly, using Bell polynomials, which we introduce in the next section.

\section{Bell Polynomials and their combinatorics}

Bell Polynomials (introduced in \cite{Bell1934}, see also sec. 3.3 in \cite{Comtet1974}) are inseparably linked with the famous Fa\'a di Bruno's formula i.e. the generalization of the chain rule (derivation of the composition of two functions) to higher derivatives. However, the formula has been published \cite{FaadiBruno1855}  30 years before Bell was born, so naturally,  it was stated without that notion (see Eq. (\ref{eq:Riordan})). We, however, use its version presented in Eq. (\ref{eq:Riordan-Bell}), where we introduced Bell polynomials. For the historical background of the Fa\'a di Bruno's formula  see \cite{Johnson2002,Craik2005}.  Fa\'a di Bruno's formula (in clasical formulation in Eq. (\ref{eq:Riordan}) and with the usage of Bell polynomials $B_{n,\,k}$ in Eq. (\ref{eq:Riordan-Bell})) states
\begin{align}\label{eq:Riordan}
\frac{d^n}{dx^n} F(G(x)) &=\sum_{\{m_r\}_n} {\frac {n!}{m_{1}!\,m_{2}!\,\cdots \,m_{n}!}}\cdot F^{(m_{1}+\cdots +m_{n})}(G(x))\cdot \prod _{j=1}^{n}\left[\frac{G^{(j)}(x)}{j!}\right]^{m_{j}}=\\
&= \sum_{k=1}^n F^{(k)}(G(x))\cdot B_{n,k}\left(G'(x),G''(x),\dots,G^{(n-k+1)}(x)\right),\label{eq:Riordan-Bell}
\end{align}
where $F$ and $G$ are analytical functions and summation in Eq. (\ref{eq:Riordan}) is taken over  integers $m_i$ such  
\begin{equation*}
1\cdot m_{1}+2\cdot m_{2}+3\cdot m_{3}+\cdots +n\cdot m_{n}=n.
\end{equation*}
Bell Polynomials $B_{nk}$ from Eq.~(\ref{eq:Riordan-Bell}) can be defined in the two equivalent ways -- combinatorically and analytically. Let us recall those two approaches in the following subsections.

\subsection{Analytic definition of Bell Polynomials}

As in the previous section, let us consider two analytic functions $F$ and $G$ such that $F(0)=0$,  which means that
\begin{equation*}
F(x)=\sum _{n=1}^{\infty }f_{n} \frac{x^{n}}{n!},\;\;\; G(x)=\sum _{n=0}^{\infty }g_{n} \frac{x^{n}}{n!}.
\end{equation*}
Then we ask about the series expansion of the composition of both functions, similarly to the consideration in Fa\'a di Bruno's formula (cf. Eqs. (\ref{eq:Riordan}, \ref{eq:Riordan-Bell}))
\begin{equation*}
G(F(x))=\sum _{n=0}^{\infty }c_{n} \frac{x^{n}}{n!}.
\end{equation*}
Let us then define Bell polynomials $B_{n,k}$ and connect them with the coefficients $c_n,\,f_n,\,g_n$ as follows
\begin{equation}\label{eq:Belldef1}
\begin{cases}
c_0=g_0,\\
c_n=\sum _{k=1}^{n}g_{k}B_{n,\,k}(f_{1},\dots ,f_{n-k+1})\;\;\mathrm{for}\;\;n>0.
\end{cases}
\end{equation}
Combining Eqs. (\ref{eq:Riordan}) and (\ref{eq:Belldef1}) one obtains the following analytical definition of Bell polynomials
\begin{equation}\label{eq:Bell-definicja}
B_{n,k}(a_{1},a_{2},\dots ,a_{n-k+1})=\sum_{\{m_r\}_{n,k}} \frac{n!}{m_{1}!m_{2}!\cdots m_{n-k+1}!}\prod _{j=1}^{n}\left(\frac{a_j}{j!}\right)^{m_{j}},
\end{equation}
where the summation is taken over all non-negative integers $\{m_r\}$ which satisfy
\begin{align}\label{eq:m}
\begin{cases}1\cdot m_{1}+2\cdot m_{2}+3\cdot m_{3}+\cdots +n\cdot m_{n}&=n,\\
 m_{1}+ m_{2}+ m_{3}+\cdots + m_{n}&=k.
 \end{cases}
\end{align}
Let us note the similarity of conditions (\ref{eq:cond}) and (\ref{eq:m}), which makes Bell polynomials a natural tool for describing Darwin-Fowler's formalism. Let us also compare definition of function $\Gamma_{\mathcal{N},\,\mathcal{K}}$ given in the Eq. (\ref{eq:Gamma}) with Bell polynomial given by the Eq. (\ref{eq:Bell-definicja}).

\subsection{Combinatorial definition of Bell Polynomials}

Now let us ask seemingly totally different and non-connected to the previous one question: what is the number of possible decompositions of a set of $n$ elements into $k$ clusters (subsets)? Additionally, we assume that every cluster of size $l$ has $a_l\geq 0$ possible configurations. Firstly, let us fix one of the set divisions. Such decomposition is described by the sequence of the non-negative integers  $\{m_r\}$ which satisfy
\begin{align*}
\begin{cases}1\cdot m_{1}+2\cdot m_{2}+3\cdot m_{3}+\cdots +n\cdot m_{n}&=n,\\
 m_{1}+ m_{2}+ m_{3}+\cdots + m_{n}&=k,
 \end{cases}
\end{align*}
where $m_\ell$ describes the number of subsets of size $\ell$. In this situation, the number of possible implementations of such a division is equal to

\begin{equation*}
    \frac{n!}{m_{1}!m_{2}!\cdots m_{n-k+1}!}\prod _{j=1}^{n}\left(\frac{a_j}{j!}\right)^{m_{j}},
\end{equation*}
which, summed over all possible partitions $\{m_r\}_{\mathcal{N},\,\mathcal{K}}$ leads to Bell polynomials. The formal proof of the equivalence between analytic and combinatorial definition of Bell Polynomials can be found in \cite{Comtet1974}, however let us describe this fact in the following example.
\begin{example}
Let us consider two examples of Bell Polynomials with $n=6$, $k=3$ and $n=6$, $k=4$. In both cases the coefficients of the polynomials count the number of a possible partition of the set into clusters of sizes given by indices of $a$: 
\begin{align*}
    B_{{6,3}}(a_{1},a_{2},a_{3},a_{4},a_5,a_6)&=\underbrace{15}_{=\binom{6}{4}}a_{1}^2a_{4}+\underbrace{60}_{=\binom{6}{3}\binom{3}{2}}a_{1}a_{2}a_{3}+\underbrace{15}_{=\frac{1}{3!}\binom{6}{2}\binom{4}{2}}a_{2}^{3},\\
B_{{6,4}}(a_{1},a_{2},a_{3},a_{4},a_5,a_6)&=\underbrace{20}_{=\binom{6}{3}} a_1^3 a_3+\underbrace{45}_{=\frac{1}{2!}\binom{6}{2}\binom{4}{2}=} a_1^2 a_2^2.
\end{align*}
\end{example}
In the following section we need the formula for the derivation of Bell polynomials
\begin{equation}\label{eq:Bellder}
    \frac{\partial B_{n,\,k}(a_1,\,a_2,\,\dots)}{\partial a_\ell}=\binom{n}{\ell}B_{n-\ell,\,k-1}(a_1,\,a_2,\,\dots),
\end{equation}
which we prove in \ref{sec:Bellder}. Let us also recall another important property of Bell Polynomials after \cite{Comtet1974} 
\begin{equation}\label{eq:Bell-relation}
B_{n,k}(abx_1, ab^2x_2, ...)=a^kb^nB_{n,k}(x_1, x_2 ,...).
\end{equation}

\subsection{Ordinary Bell polynomials}
The Bell polynomials $B_{n,k}$ discussed in the previous sections are sometimes called exponential polynomials (see \cite{Comtet1974}). Let us consider a minor modification of them $\widehat{B}_{n,k}$, which for the sake of distinction, are called ordinary Bell polynomials (see Eq. [3o] in~\cite{Comtet1974}).
\begin{equation}\label{eq:OBP}
    \widehat{B}_{n,k}(a_{1},a_{2},\dots )={\frac {k!}{n!}}B_{n,k}(1!a_{1},2!a_{2},\dots)=\sum_{\{m_r\}_{n,k}} \frac{k!}{m_{1}!m_{2}!\cdots m_{n-k+1}!}\prod _{j=1}^{n}a_j^{m_{j}},
\end{equation}
As we will see in the following sections, ordinary polynomials are more natural for describing the Darwin-Fowler procedure, rather than exponential ones.

\subsection{Bell transformation and combinatorial numbers}
Bell polynomials also allow one to conveniently express known combinatorial numbers in a compact form. There are several versions of the so-called Bell transform (see \cite{Birmayer2019,Lushny}), but we will focus on expressing Lah and Stirling numbers by exponential Bell polynomials (see \cite{Comtet1974,Charalambides2002}). As we will see in the following sections, these numbers are closely related to the distributions in the Darwin-Fowler approach.

\begin{itemize}
    \item Unsigned Lah numbers (see A105278 in \cite{oeis} and Eq. [3h] in \cite{Comtet1974}) 
\begin{equation}\label{eq:Lah}
    L(n,k)=B_{n,k}(1!,2!,\dots)=\binom{n-1}{k-1}\frac{n!}{k!}.
\end{equation}
\item Unsigned Stirling numbers of the first kind (see A008275 in \cite{oeis} and Eq. [3i] in \cite{Comtet1974}) 
\begin{equation}\label{eq:S1}
    |S_1(n,k)|=B_{n,k}(0!,1!,2!,\dots).
\end{equation}
\item Stirling numbers of the second kind (see A008277 in \cite{oeis} and Eq. [3g] in  \cite{Comtet1974}) 
\begin{equation}\label{eq:S2}
    S_2(n,k)=B_{n,k}(1,1,1,\dots).
\end{equation}

\end{itemize}

\section{Bell Polynomial approach to  Darwin-Fowler model}
Using Bell polynomials (both exponential and ordinary) enables to  transform the expression for $\Gamma_{\mathcal{N},\,\mathcal{K}}$ given by Eq. (\ref{eq:Gamma}) as follows
\begin{equation}\label{eq:GammaB}
	\Gamma(\omega_1,\,\omega_2,\,\dots)=\frac{\mathcal{K}!}{\mathcal{N}!}B_{\mathcal{N},\,\mathcal{K}}(1!\omega_1,\,2!\omega_2,\,\dots)=\widehat{B}_{\mathcal{N},\,\mathcal{K}}(\omega_1,\,\omega_2,\,\dots).
\end{equation}
Thus, the expression for a $n^\star_{\mathcal{N},\,\mathcal{K}}$ (see Eqs. (\ref{eq:wartoscocz}) and (\ref{eq:OBP})) which we are looking for can be further simplified 
\begin{equation}\label{n_sr_Belle}
n^\star_{\mathcal{N},\,\mathcal{K}}(\ell)=\frac{\mathcal{N}!}{(\mathcal{N}-\ell)!} \frac{\omega_\ell B_{\mathcal{N}-\ell, \mathcal{K}-1}(1!\omega_1, 2!\omega_2, \cdots)}{B_{\mathcal{N},\mathcal{K}}(1!\omega_1, 2!\omega_2, \cdots)}=\mathcal{K}\underbrace{\frac{\omega_\ell \widehat{B}_{\mathcal{N}-\ell, \mathcal{K}-1}(\omega_1, \omega_2, \cdots)}{\widehat{B}_{\mathcal{N},\mathcal{K}}(\omega_1, \omega_2,\cdots)}}_{=\mathcal{P}_{\mathcal{N},\,\mathcal{K}}(\ell)},
\end{equation}
which is the exact result for finite $\mathcal{K}$ and $\mathcal{N}$. As illustrations of this main result let us consider specific forms of degeneration in the following sections.

\section{Special cases of the degeneracy}

\subsection{Constant degeneracy}
Firstly let us assume that there is no degeneracy, i.e. $ \omega_\ell=\omega=\mathrm{constans}$, which changes Eq. (\ref{n_sr_Belle}) to the following form (see Eq. (\ref{eq:Bell-relation}))
\begin{equation*}
	n^\star_{\mathcal{N},\mathcal{K}}(\ell)=\frac{\mathcal{N}!}{\left(\mathcal{N}-\ell\right)!}\frac{B_{\mathcal{N}-\ell,\,\mathcal{K}-1}(1!,\,2!,\,3!,\,\dots)}{B_{\mathcal{N},\,\mathcal{K}}(1!,\,2!,\,3!,\,\dots)}=\frac{\mathcal{N}!}{\left(\mathcal{N}-\ell\right)!}\frac{L(\mathcal{N}-\ell,\,\mathcal{K}-1)}{L(\mathcal{N},\,\mathcal{K})},
\end{equation*}
where we can spot  Lah numbers  (see Eq. (\ref{eq:Lah})), which can be further simplified
\begin{align}\nonumber
	n^\star_{\mathcal{N},\mathcal{K}}(\ell)&=\frac{\mathcal{N}!}{\left(\mathcal{N}-\ell\right)!}\frac{\binom{\mathcal{N}-\ell-1}{\mathcal{K}-2}\frac{\left(\mathcal{N}-\ell\right)!}{\left(\mathcal{K}-1\right)!}}{\binom{\mathcal{N}-1}{\mathcal{K}-1}\frac{\mathcal{N}!}{\mathcal{K}!}}=\mathcal{K}\frac{\binom{\mathcal{N}-\ell-1}{\mathcal{K}-2}}{\binom{\mathcal{N}-1}{\mathcal{K}-1}}=\\&=\mathcal{K}\underbrace{\left(\mathcal{K}-1\right)\frac{\left(\mathcal{N}-\ell-1\right)!\left(\mathcal{N}-\mathcal{K}\right)!}{\left(\mathcal{N}-1\right)!\left(\mathcal{N}-\mathcal{K}-\ell+1\right)!}}_{\mathcal{P}_{\mathcal{N},\mathcal{K}}(\ell)}.\label{eq:kanonicznyLah}
\end{align}
Typically we would look for the asymptotic form of Eq. (\ref{eq:kanonicznyLah}) using the steepest descent method, which is described in \ref{sec:steepestdescend}. However, we will use the compact form of Lah numbers and apply the Stirling's approximation in  $\spadesuit$ and obtain
\begin{align*}
\mathcal{P}_{\mathcal{N},\mathcal{K}}=&\left(\mathcal{K}-1\right)\frac{\left(\mathcal{N}-\ell-1\right)!}{\left(\mathcal{N}-1\right)!}\frac{\left(\mathcal{N}-\mathcal{K}\right)!}{\left(\mathcal{N}-\mathcal{K}-\ell+1\right)!}\stackrel{\spadesuit}{\approx}\\
\stackrel{\spadesuit}{\approx}& \frac{\left(\mathcal{K}-1\right)\left(\mathcal{N}-\ell-1\right)^{\mathcal{N}-\ell-1}e^{-\left(\mathcal{N}-\ell-1\right)}}{\left(\mathcal{N}-1\right)^{\mathcal{N}-1}e^{-\left(\mathcal{N}-1\right)}}\frac{\left(\mathcal{N}-\mathcal{K}\right)^{\mathcal{N}-\mathcal{K}}e^{-\left(\mathcal{N}-\mathcal{K}\right)}}{\left(\mathcal{N}-\mathcal{K}-\ell+1\right)^{\mathcal{N}-\mathcal{K}-\ell+1}e^{-\left(\mathcal{N}-\mathcal{K}-\ell+1\right)}}=\\
=& \frac{\left(\mathcal{K}-1\right) e\left(\mathcal{N}-\mathcal{K}\right)^{\ell-1}}{\left(\mathcal{N}-\ell-1\right)^{\ell}}\left(1-\frac{\ell}{\mathcal{N}-1}\right)^{\mathcal{N}-1} \left(1+\frac{\ell-1}{\mathcal{N}-\mathcal{K}-\ell+1}\right)^{\mathcal{N}-\mathcal{K}-\ell+1} =\\
=& e\frac{\mathcal{K}-1}{\mathcal{N}-\mathcal{K}}\left( \frac{\mathcal{N}-\mathcal{K}}{\mathcal{N}-\ell-1}\right)^\ell \left(1-\frac{\ell}{\mathcal{N}-1}\right)^{\mathcal{N}-1} \left(1+\frac{\ell-1}{\mathcal{N}-\mathcal{K}-\ell+1}\right)^{\mathcal{N}-\mathcal{K}-\ell+1}.
\end{align*}
Reasonably assuming that the average energy $U$ of the system is constant even in the limit of large $\mathcal{K}$ i.e.   
$\mathcal{N}=U \mathcal{K}$ one gets final form of the energy distribution as
\begin{align*}
\mathcal{P}_{\mathcal{N},\,\mathcal{K}}(\ell)\approx& e \underbrace{\frac{\mathcal{K}-1}{(U-1)\mathcal{K}}}_{\xrightarrow{\mathcal{K}\rightarrow \infty} (U-1)^{-1}}\underbrace{\left( \frac{(U-1)\mathcal{K}}{U\mathcal{K}-\ell-1}\right)^\ell}_{\xrightarrow{\mathcal{K}\rightarrow \infty} \left(\frac{U-1}{U}\right)^\ell}\underbrace{\left(1-\frac{\ell}{U\mathcal{K}-1}\right)^{U\mathcal{K}-1}}_{\xrightarrow{\mathcal{K}\rightarrow \infty} e^{-\ell}}\times\\ & \times\underbrace{\left(1+\frac{\ell-1}{(U-1)\mathcal{K}-\ell+1}\right)^{(U-1)\mathcal{K}-\ell+1}}_{\xrightarrow{\mathcal{K}\rightarrow \infty} e^{\ell-1}},
\end{align*}
which results in the expected formula for the distribution
\begin{equation}\label{eq:Pkanon}
\mathcal{P}_\infty(\ell)=\frac{e^{\beta \varepsilon \ell}}{Z},
\end{equation}
where $U=(1-e^{\beta \varepsilon})^{-1}$ and $Z=U-1$.

\subsection{Harmonic oscillators}
For the one-dimensional harmonic oscillator the weight of energy $\varepsilon_\ell$ is equal to
\begin{equation*}
\omega_\ell=(\ell+1),
\end{equation*} 
which implies the following form of the energy distribution from Eq. (\ref{n_sr_Belle})
\begin{equation}\label{eq:HO1}
\mathcal{P}_{\mathcal{N},\,\mathcal{K}}(\ell)=(\ell+1)\frac{ \widehat{B}_{\mathcal{N}-\ell, \mathcal{K}-1}(2, 3, \dots)}{\widehat{B}_{\mathcal{N},\mathcal{K}}(2, 3,\dots)}.
\end{equation}
Eq. (\ref{eq:HO1}) can be further generalized for the $D$-dimensional harmonic oscillator, because degenerations are then given as
\begin{equation*}
\omega_\ell=(\ell+1)(\ell+2)\dots(\ell+D)=:(\ell+1)^{(D)},
\end{equation*} 
where $(\ell+1)^{(D)}$ denotes the rising factorial for brevity. With such degeneracies one face the problem of  determination the values of  Bell polynomials $B_{n,k}(2^{(D)},3^{(D)},\dots)$, which can be done with the following formula
\begin{equation}\label{Bell_Stirling}
B_{n,k}(2^{(D)},3^{(D)},\dots)=\left[ (D-1)!\right]^k \sum_{r=k}^n |S_1(n,r)|S_2(r,k)D^r, 
\end{equation}
where $S_1, S_2$ are the Stirling numbers of the first and second kind respectivelly, see Eqs. (\ref{eq:S1}) and (\ref{eq:S2}). Let us note that  Eq. (\ref{Bell_Stirling}) follows from Eq. (8.50) in \cite{Charalambides2002} and allows one to obtain the following final formula for the most probable configuration for $D$-dimensional harmonic oscillator
\begin{equation}\label{nsroscylatory}
n^\star_{\mathcal{N},\mathcal{K}}=\frac{\mathcal{N}!}{(\mathcal{N}-\ell)!}\frac{(\ell+1)^{(D)}}{(D-1)!}\frac{\sum_{r=\mathcal{K}-1}^{\mathcal{N}-\ell}{|S_1(\mathcal{N}-\ell,r)|S_2(r,\mathcal{K}-1)D^r}}{\sum_{r=\mathcal{K}}^\mathcal{N} |S_1(\mathcal{N}, r)| S_2(r,\mathcal{K})D^r}.
\end{equation}

As we can see, the above Eq. (\ref{nsroscylatory}) combines the energy distribution for a harmonic oscillator with the Stirling numbers.

\section{Acknowledgement}
We would like to thank Agata Fronczak for stimulating discussions (and pointing out our mistakes at the early stage of the work). GS work has been supported by the National Science Centre of Poland (Narodowe Centrum Nauki, NCN) under grant no. 2015/18/E/ST2/00560.

\appendix

\section{Steepest descend approach to Darwin-Fowler}\label{sec:steepestdescend}
As we mentioned previously, typically in Darwin-Fowler approach one is  only interested in the result in the thermodynamical limit $\mathcal{K}\rightarrow \infty$ with $\mathcal{N}=U\mathcal{K}$. Keeping the above in mind let us define a generating function for $\Gamma_{\mathcal{N},\mathcal{K}}$ in the following way 
\begin{equation}
\label{generating_fun}
G_\mathcal{K}(z)=\sum_{U=0}^{\infty}z^{\mathcal{K} U}\Gamma_{\mathcal{K} U,\mathcal{K}},
\end{equation} 
It is easy to see (compare Eq. (\ref{eq:Gamma}) or more detailed discussion in \cite{Huang1978,Pathria2011}) that due to the multinomial theorem this generating function simplifies to 
\begin{equation}
\label{generating_sum_sym}
G_\mathcal{K}(z)=(\omega_1z^{\varepsilon_1}+ \omega_2z^{\varepsilon_2}+ \dots)^\mathcal{K} = \left[ g(z) \right]^\mathcal{K}.
\end{equation}
Let us now focus on the specific case and assume (after Huang \cite{Huang1978}) that degeneracies are constant i.e. $\omega_\ell=1$, and $\varepsilon=1$ which simplifies the function as follows 
\begin{align}\label{part_function}
g(z)= 1+z+ z^2+ z^{3} +\dots=\frac{1}{1-z}. 
\end{align}
From Eq. (\ref{part_function}) one see that  $\Gamma_{\mathcal{K}U,\mathcal{K}}$ is the coefficient of $z^{\mathcal{K}U}$ in the expansion of $G_\mathcal{K}(z)$ in powers of $z$,  hence 
\begin{equation}
\label{Gamma_integer}
\Gamma_{\mathcal{K}U,\mathcal{K}} = \frac{1}{2\pi i}\oint  \frac{[g(z)]^\mathcal{K}}{z^{\mathcal{K}U+1}}dz.
\end{equation}
For real positive $z$ function $g(z)$  monotonically increases  with a radius of convergence $z=R$. The function $1/z^{\mathcal{K}U+1}$ is a monotically descreasing function of real  positive $z$. Hence the function $[f(z)]^\mathcal{K}/z^{\mathcal{K}U+1}$ has a minimum at $z=x_0$ for $z\in[0,R]$.  In addition, functions $f(z)$ and $1/z^{\mathcal{K}U+1}$ are analytic, therefore the integrand $I(z)=[f(z)]^\mathcal{K}/z^{\mathcal{K}U+1}$ is also analytic and satisfies the Cauchy-Riemann equation
\begin{align*}
\left( \frac{\partial^2}{\partial x^2}+ \frac{\partial^2}{\partial y^2}\right)=0,
\end{align*}
hence 
\begin{equation}
\label{cauchy}
\left( \frac{\partial I}{\partial z}\right)_{z=x_0}=0, \quad \left( \frac{\partial^2 I}{\partial x^2}\right)_{z=x_0}> 0, \quad  \left( \frac{\partial^2 I}{\partial y^2}\right)_{z=x_0}> 0,
\end{equation}
where $z=xi+y$. One can see that $x_0$ is a saddle point. Thus let us define $u(z)$ as
\begin{align}
I(z)=e^{\mathcal{K}u(z)},\;\;\;u(z)=\ln g(z)-(\mathcal{K}U+1)\ln z .
\end{align}
Taking (\ref{cauchy}) into account, one can obtain
\begin{align*}
\frac{\partial^2 I}{\partial x^2} \xrightarrow[]{\mathcal{K}\rightarrow \infty}\infty,\;\;\;\;
\frac{\partial^2 I}{\partial y^2} \xrightarrow[]{\mathcal{K}\rightarrow \infty}\infty.
\end{align*}
Therefore the saddle point touches an infinitely sharp peak and an infinitely steep valley in the limits as $\mathcal{K}\rightarrow \infty$. 
If we choose the contour of integration to be a circle centered in $z=0$ with radius $x_0$ the main part of integral  comes from the neighrhood  of $x_0$. Thus to compute the integral one can extend the intergrand around $z=x_0$, hence
\begin{equation}
\label{int_result}
\Gamma_{\mathcal{K}U,\mathcal{K}}=\frac{1}{2\pi i}\oint e^{\mathcal{K}u(z)}dz\approx e^{\mathcal{K}u(x_0)} \frac{1}{2\pi} \int_{-x_0}^{x_0} dye^{-1/2 \mathcal{K}u''(x_0)y^2}\approx \frac{e^{\mathcal{K}u(x_0)}}{\sqrt{2\pi \mathcal{K} u''(x_0)}}.
\end{equation}
From Eqs. (\ref{eq:wartoscocz}) and (\ref{int_result}), the average number $n^\star(\ell)$ of systems in the degenerate state of energy $\varepsilon_l\varepsilon \ell$ is thus given by the expression
\begin{equation}\label{n_sr}
n^\star(\ell)\propto \exp (\beta\varepsilon \ell),
\end{equation}
which is consistent with the result from Bell polynomial approach (see Eq. (\ref{eq:Pkanon})).

\section{Proof of Eq. \ref{eq:Bellder}}\label{sec:Bellder}
We want to prove that the derivative of Bell polynomial follows Eq. (\ref{eq:Bellder}) i.e.
\begin{equation*}
    \frac{\partial B_{n,\,k}(a_1,\,a_2,\,\dots)}{\partial a_\ell}=\binom{n}{\ell}B_{n-\ell,\,k-1}(a_1,\,a_2,\,\dots).
\end{equation*}
Let us start with the left-hand side of the original equation for $n>\ell$ and $k>1$, othervise derivative is equal to zero
\begin{equation*}
    \frac{\partial B_{n,\,k}(a_1,\,a_2,\,\dots)}{\partial a_\ell}=\sum_{\{m_r\}_{n,k}} \frac{n!}{m_{1}!\cdots m_{n-k+1}!}\left(\frac{1}{\ell!}\right)^{m_\ell}\frac{\partial a_\ell^{m_\ell}}{\partial a_\ell}\prod _{j\neq\ell}\left(\frac{a_j}{j!}\right)^{m_{j}}=\left(\spadesuit\right),
\end{equation*}
which can be further transformed as follows due to the elementary differentiation
\begin{align*}
    \left(\spadesuit\right)=&\sum_{\substack{\{m_r\}_{n,k},\\m_\ell\neq 0}} \frac{n!\left(\ell!\right)^{-m_\ell}\prod _{j\neq\ell}\left(\frac{a_j}{j!}\right)^{m_{j}}}{m_{1}!\cdots m_{n-k+1}!}\frac{\partial a_\ell^{m_\ell}}{\partial a_\ell}+\underbrace{\sum_{\substack{\{m_r\}_{n,k},\\m_\ell= 0}} \frac{n!\left(\ell!\right)^{-m_\ell}\prod _{j\neq\ell}\left(\frac{a_j}{j!}\right)^{m_{j}}}{m_{1}!\cdots m_{n-k+1}!}\frac{\partial a_\ell^{m_\ell}}{\partial a_\ell}}_{=0}=\\
    =&\frac{1}{\ell!}\sum_{\substack{\{m_r\}_{n,k},\\m_\ell\neq 0}} \frac{n!}{m_{1}!\cdots(m_\ell-1)!\cdots m_{n-k+1}!}\left(\frac{a_\ell}{\ell!}\right)^{m_\ell-1}\prod _{j\neq\ell}\left(\frac{a_j}{j!}\right)^{m_{j}}=(\clubsuit).
\end{align*}
Let us note that thanks to the condition of the summation given by Eq. (\ref{eq:m}) one can simplify the above into the form
\begin{equation*}
    (\clubsuit)=\frac{1}{\ell!}\sum_{\{m_r\}_{n-\ell,k-1}} \frac{n!}{m_{1}!\cdots(m_\ell)!\cdots m_{n-k+1}!}\prod _{j}\left(\frac{a_j}{j!}\right)^{m_{j}}=\frac{n!}{\ell!(n-\ell)!} B_{n-\ell,\,k-1}(a_1,\,a_2,\,\dots),
\end{equation*}
which ends the proof.

\end{document}